\documentstyle[aps,preprint]{revtex}
\begin{document}
\draft
%\preprint{SOGANG-HEP 217/97}
\preprint{\vbox{ \hbox{SOGANG-HEP 217/97} \hbox{hep-th/9704115} }}
\title{Quantization of dilaton cosmology in two dimensions}
\author{Won T. Kim\footnote{electronic address:wtkim@ccs.sogang.ac.kr}
  and Myung Seok Yoon\footnote{electronic address:younms@physics3.sogang.ac.kr}}
\address
{Department of Physics and Basic Science Research Institute, \\ 
Sogang University, Seoul 121-742, Korea }
\date{April 1997}
\maketitle
\bigskip
\begin{abstract}
 The generalized kinetic term of a dilaton gives the classical 
superinflation without recourse to any potential, and the quantum 
version of the dilaton gravity exhibits the finite curvature and 
graceful exit. For $p=2$ case, the model corresponds to the RST 
quantization of the s-wave sector of the four-dimensional Einstein 
cosmology. Further, the de Sitter universe is realized for $p=8$
and the smooth transition to the Minkowski space-time is possible. 
Even in the accelerating contraction case of the universe for $-4<p<0$,
the curvature singularity does not appear in a certain branch.
\end{abstract}
\bigskip

\newpage
\begin{center}
{\bf I. Introduction}
\end{center}
%\section{Introduction}\label{sec:intro}

  There has been much interest in the cosmological problems in the
low-energy string theory. Recently, it has been shown that the scale
factor duality~\cite{vst} of the string theory motivated the inflation
scenario~\cite{gv}. The natural graceful exit~\cite{bv} of
superinflation to the decelerating expansion phase has not been
realized in the classical sting cosmology. The classical curvature 
singularity separates two classical solutions corresponding to
accelerating expansion and decelerating expansion, respectively.
However, the finite quantum transition between two cosmological phases
is possible~\cite{gmv}. The explicit resolution of the graceful exit 
problem  was suggested by considering the quantum back reaction of 
the geometry in the two-dimensional string theory~\cite{rey} and 
the one-loop superstring cosmology and the dilaton cosmology~\cite{em}. 
In fact, the curvature singularity can be mild by the quantum back 
reaction of the geometry of the black hole in the spherical symmetric 
part of the Einstein gravity~\cite{ks}. So it seems to be very
plausible to reconcile the singularity problems with the help of
the quantum back reaction in the gravity theory.

 On the other hand, the Callan-Giddings-Harvey-Strominger(CGHS)~\cite{cghs} 
and the Russo-Susskind-Thorlacius (RST)~\cite{rst} models have been 
extensively studied in order to study the final state of the black 
hole and the various aspects of the quantum back reaction~\cite{wl}. 
The exactly soluble gravity models are crucial to determine the
correct evolution of the geometry~\cite{mr}. At the same time, 
they are more amenable to quantum treatments than their 
four-dimensional counterparts.
  
 In this paper, we shall study the exactly soluble dilaton-gravity model 
which has a generalized coefficient in the kinetic term of a dilaton.
This model contains the s-wave sector of Einstein gravity for $p=2$. 
Further, for $p<0$, the kinetic part has the right sign which
corresponds to the two-dimensional version of the Brans-Dicke theory. 
However, the range of coefficient is restricted to $-4< p \leq 8$ 
in the first branch and $p>0$ in the second branch to obtain the
finite curvature scalar in the quantized theory. In this range, 
the duality-related solutions are possible only for the positive 
definite $p$. For the special case of $p=8$, the classical 
de Sitter universe is realized. The quantum correction modifies 
the geometry and the de Sitter universe appears only asymptotically. 
In sect.~II, the motivation is presented in terms of the spherical 
symmetric reduction of the Einstein gravity. The detailed analysis 
of our model is given in sect.~III. The case of $p=8$ is briefly 
analyzed in sect.~IV. Finally the discussion is given in sect.~V.

\begin{center}
{\bf II. Motivation}
\end{center}
%\section{Motivation}\label{sec:motiv}

 The most general spherically symmetric metric can expressed 
in the form of
\begin{equation}
ds^2=ds^2_{(2)} + \frac{1}{\lambda^2} e^{-2\phi}\,d\Omega^2 ,
\end{equation}
where $2\lambda^2=\frac{\pi}{G_N}$ and $G_N$ is a Newton's 
constant~\cite{str-lec}. The spherical symmetric reduction of the 
Einstein-Hilbert action leads to a two-dimensional dilaton gravity,
\begin{equation}
\label{ca}
S_{{\rm cl}} = \frac{1}{2 \pi} \int\/d^{2}x\sqrt{-g}\left\{
	e^{-2\phi}\left[ R + 2(\nabla\phi)^2 \right] + 2\lambda^2 \right\},
\end{equation}
where the cosmological constant in two dimensions is inversely
proportional to the Newton constant. For simplicity, let us consider
the very strong coupling limit, i.e., the vanishing cosmological
constant. The classical equations of motion yield cosmological
solutions with two classical branches in the conformal gauge,
\begin{equation}
g_{\pm\mp} = - \frac12 e^{2\rho},\qquad g_{\pm\pm} = 0.
\end{equation}
The first branch corresponding to the accelerating
expansion(pre-big-bang phase) is 
\begin{eqnarray}
\rho(t) &=& \frac{1}{2}\phi(t) - \ln A , \\
e^{-2\phi(t)} &=& \Sigma t ,
\end{eqnarray}
where $A$ and $\Sigma$ are integration constants. 
In the comoving time, the scale factor and the curvature are given by
\begin{equation}
\label{scale}
a(\tau) \sim (-\tau)^{-\frac{1}{3}}, \qquad R(\tau) \sim (-\tau)^{-2} ,
\end{equation}
where the range of comoving time is $-\infty <\tau <0 $. This branch 
exhibits the superinflation behavior and the curvature scalar diverges
at $\tau =0$. The other branch (post-big-bang phase) is flat space-time,
\begin{equation}\label{cl:+}
a(\tau) \sim \tau , \qquad R(\tau) = 0,
\end{equation}
where $0<\tau < +\infty$. The above two branches are not connected 
smoothly because of the curvature singularity at $\tau=0$. This
feature already has been shown by Rey~\cite{rey} in the CGHS model 
with vanishing cosmological constant. The only difference comes from 
the power of scale factor, where $a(\tau) \sim (-\tau)^{-1}$ 
in the classical low energy string theory. As a matter of fact, 
the cosmological constant is generically nonzero. So it is unclear 
whether the classical superinflation of the first branch in our model 
is possible or not even in the presence of the cosmological constant. 
In the quantized theory which will be shown in the next section, 
however, the effective cosmological constant can be set to zero. 

The two-dimensional dilaton gravity model is analyzed in~\cite{gv2d} 
by generalizing the central charge of the conformal anomaly and
the coefficient of the local counter term. In the next section, we 
shall consider the two-dimensional dilaton-gravity model with
the generalized dilaton kinetic term.
 
\begin{center}
{\bf III. Dilaton Cosmology} 
\end{center}
%\section{Dilaton Cosmology ($-4<p<8$)}\label{sec:dc}

 Let us now start with the effective action given by
\begin{equation}
\label{total}
S=S_{{\rm{cl}}} + S_{{\rm{qt}}},
\end{equation}
\begin{eqnarray}
S_{{\rm{cl}}}& =& \frac{1}{2 \pi} \int\/d^{2}x\sqrt{-g}\left\{
	e^{-2\phi}\left[ R + p(\nabla\phi)^2 \right] + 2\lambda^2
	- \frac12 e^{-2a\phi}\sum_i (\nabla f_i)^2  \right\}, \\
S_{{\rm{qt}}}& =& \frac{1}{2\pi}\int\/d^2 x\sqrt{-g}\left\{ 
	-\frac{\kappa}{4}R\frac{1}{\Box}R - \frac{\gamma}{2}\phi R 
	- \Lambda \right\},
\end{eqnarray}
where $i=1,2,\cdots N$, and $a=0$ for the scalar fields and $a=1$ for 
the bosonized fields of s-wave fermions in the four-dimensional point
of view. We generalize the coefficient of the kinetic part of
dilaton. The conformal anomaly coefficient $\kappa$ is not positive 
definite and $\gamma=\frac{p\kappa}{4}$ is determined to solve exactly
up to the quantum level. Further, the local ambiguity of conformal 
anomaly $\Lambda$ is chosen to cancel the classical cosmological 
constant, $\Lambda=2\lambda^2$. Hence we naturally obtain the  exactly
solvable model.

In the conformal gauge, the gauge fixed action is written as
\begin{equation}\label{gauge-action}
S = \frac{1}{\pi}\int\/d^2 x \left\{ 
	e^{-2\phi}\left[ 2\partial_+\partial_-\rho
	- p\partial_+\phi\partial_-\phi \right] 
	- \kappa\partial_+\rho\partial_-\rho 
	- \frac{p\kappa}{4}\phi\partial_+\partial_-\rho 
	+ \frac12 e^{-2a\phi}\sum_i \partial_+f_i\partial_-f_i          \right\}.
\end{equation}
The equations of motion with respect to the metric, dilaton, and
matters are given by
\begin{eqnarray}
& &T_{+-} = e^{-2\phi}\left[2\partial_+\partial_-\phi -
	4\partial_+\phi\partial_-\phi\right] +
	\frac{p\kappa}{8}\partial_+\partial_-\phi -
	\kappa\partial_+\partial_-\rho = 0, \label{+-} \\
& &e^{-2\phi}\left[\partial_+\partial_-\rho +
	\frac{p}{2}\left(\partial_+\phi\partial_-\phi -
	\partial_+\partial_-\phi \right)\right] +
	\frac{p\kappa}{8}\partial_+\partial_-\rho  +
	\frac{1}{4}ae^{-2a\phi}\sum_i \partial_+f_i\partial_-f_i = 0, 
	\label{dilaton} \\
& &\partial_+\partial_-f_i - a(\partial_+\phi\partial_-f_i +
	\partial_-\phi\partial_+f_i) = 0, 
\end{eqnarray}
with two constraints implemented by the conformal gauge fixing,
\begin{eqnarray}
\label{ppmm}
T_{\pm\pm} &=& e^{-2\phi}\left[ (4-p)(\partial_{\pm}\phi)^2 -
2\partial_\pm^2\phi + 4\partial_\pm\rho\partial_\pm\phi\right] +
\kappa\left[\partial_\pm^2\rho - (\partial_\pm\rho)^2\right] \nonumber\\
	& &- \frac{p\kappa}{8}\left[\partial_\pm^2\phi -
2\partial_\pm\rho\partial_\pm\phi\right] +
\frac{1}{2}e^{-2a\phi}\sum_i(\partial_\pm f_i)^2 - \kappa t_\pm \nonumber \\
	&= & 0.
\end{eqnarray}
The $t_\pm$ reflects the nonlocality of the induced gravity of the
conformal anomaly. Without the classical matter, $f_i =0$,
defining new fields as follows \cite{bc,dea},
\begin{eqnarray}
\Omega &=& \frac{p\kappa}{8}\phi + e^{-2\phi} \label{Omega},\\
\chi &=& \kappa\rho - \frac{p\kappa}{8}\phi+e^{-2\phi} \label{chi},
\end{eqnarray}
the gauge fixed action~(\ref{gauge-action}) is obtained in the simple form
\begin{equation}
S = \frac{1}{\pi} \int\/d^2 x \left[
	\frac{1}{\kappa}\partial_+\Omega\partial_-\Omega
	-\frac{1}{\kappa}\partial_+\chi\partial_-\chi \right]
\end{equation}
and Eqs. (\ref{+-}), (\ref{dilaton}) and (\ref{ppmm}) yield
\begin{eqnarray}
0 &=& \partial_+\partial_-\chi =\partial_+\partial_-\Omega, \label{eq:chi}\\ 
T_{\pm\pm} &=& \frac{1}{\kappa}(\partial_{\pm}\Omega)^2
		- \frac{1}{\kappa}(\partial_{\pm} \chi)^2 
		+ \partial_{\pm}^2\chi -\kappa t_{\pm} \label{eq:constr.},
\end{eqnarray}
respectively. In the homogeneous space, all fields depend only 
on the conformal time. The solutions are easily obtained as
\begin{eqnarray}
\Omega &=& \frac{p\kappa}{8}\phi + e^{-2\phi} 
		= \Omega_0 t + B ,\label{sol:Omega}\\
\chi &=& \kappa\rho - \frac{p\kappa}{8}\phi + e^{-2\phi} 
		= \chi_0 t+A \label{sol:chi},
\end{eqnarray}
with the following constraint,
\begin{equation}
\kappa t_{\pm}-\frac{1}{4\kappa}(\Omega_0 - \chi_0)(\Omega_0+\chi_0)=0,
\end{equation}
where $\Omega_0$, $\chi_0$, $A$, and $B$ are constants.
The function $t_{\pm}$ depends on the quantum matters, and it is
naturally fixed $t_{\pm}=0$ from the boundary condition for the flat 
Minkowski space time. Hereafter, we set $A=B=0$ without loss of 
generality, which has no essential effect.

The first quantum branch defined by $\Omega_0 =\chi_0$ ($-4<p<8$) 
is given by
\begin{eqnarray}
e^{-\frac{8}{p}\rho} + \frac{\kappa}{2}\rho &=& \chi_0 t, \\
e^{-2\phi} + \frac{p\kappa}{8}\phi &=& \chi_0 t,
\end{eqnarray}
where we assume $\chi_0 < 0$ so as to obtain the cosmological
expansion solution. For a convenience, by using the following relation
\begin{eqnarray}
\ddot{\rho}(t) &=& \frac{p}{4} \ddot{\phi}(t) \nonumber \\
	&=& \frac{p}{8}\chi_0^2 e^{-2\phi}\left[ e^{-2\phi} -
		\frac{p\kappa}{16}\right]^{-3} \label{ddrho},
\end{eqnarray}
which can be calculated from Eqs. (\ref{Omega}), (\ref{chi}), and 
(\ref{eq:chi}), the scalar curvature can be expressed in the form of
\begin{eqnarray}
R &=& 2e^{-2\rho}\ddot{\rho}(t) \nonumber \\
  &=& \frac{p}{4} \chi_0^2 e^{-2\rho} e^{-2\phi} \left[ 
	e^{-2\phi} - \frac{p\kappa}{16} \right]^{-3}. \label{R}
\end{eqnarray}
Note that the curvature scalar is nonsingular when $p\kappa <0$ which
is our another restriction. We now obtain the extremum condition of
curvature scalar, 
\begin{eqnarray}
\label{extremum}
\dot{R}(t) &=& \frac{p}{16}(p-8) \left(e^{-2\phi}\right)^{1+\frac{p}{4}}
		\left(e^{-2\phi}-\frac{p\kappa}{16}\right)^{-5} \left[ 
		e^{-2\phi} - \left(\frac{p+4}{p-8}\right)\frac{p\kappa}{16}
		\right] \nonumber \\ 
           &=& 0. 
\end{eqnarray}
From Eq.~(\ref{extremum}), the extremum condition is
$e^{-2\phi} - \left(\frac{p+4}{p-8}\right)\frac{p\kappa}{16}=0$ 
and the reality condition of the dilaton field gives the assumed 
restriction: $-4<p<8$.

In fact, the curvature has a maximum value for $p>0$ (a minimum
value for $p<0$) at a certain conformal time.
In terms of the comoving time defined by $\tau = \int dt\,e^{\rho(t)}$,
the metric is recast in the form
at the far past $t \rightarrow -\infty$ $(\tau \rightarrow -\infty)$, 
\begin{eqnarray}\label{1:ds1}
ds^2 &\rightarrow& -(\chi_0 t)^{-\frac{p}{4}}(dt^2 - dx^2)
              \nonumber \\
	&=& - d\tau^2 + \left[ \chi_0(1-\frac{p}{8})\tau 
	\right]^{-\frac{2p}{8-p}} dx^2 , 
\end{eqnarray}
through $\tau \approx \frac{8}{8-p}\chi_0^{-1}\left(
\chi_0 t\right)^{1-\frac{p}{8}}$. The curvature scalar is 
asymptotically flat
\begin{equation}\label{1:R1}
R \rightarrow \frac{p}{4}\left[-(1-\frac{p}{8})\tau\right]^{-2}
\end{equation}
and increases as time goes on and reaches the maximum value.
The time evolutions of the scale factor are
\begin{eqnarray}
\dot{a}(\tau) &\approx& -\frac{p}{8}\chi_0 \left[
	\chi_0(1-\frac{p}{8})\tau\right]^{-\frac{8}{8-p}} \sim sgn(p), \\
\ddot{a}(\tau) &\approx& \frac{p}{8}\chi_0^2 \left[
      \chi_0(1-\frac{p}{8})\tau\right]^{-\frac{16-p}{8-p}} \sim sgn(p), 
\end{eqnarray}
and the expansion or contraction depends on the sign of $p$. If $p>0$, 
the behavior of the space-time is accelerating expansion while it is 
accelerating contraction for $p<0$. 

On the other hand, in the far future $t \rightarrow +\infty$, 
the metric and the curvature scalar are written as
\begin{eqnarray}
ds^2 &\rightarrow& - e^{\frac{4}{\kappa}\chi_0 t}(dt^2 - dx^2)\nonumber \\
	&=& -d\tau^2 + \left(\frac{2}{\kappa}\chi_0 \tau\right)^2 dx^2,\\
R(\tau) &\rightarrow& \frac{p}{4}\left( -\frac{16}{p\kappa}\right)^3
	\chi_0^2\left(
	\frac{2}{\kappa}\chi_0\tau\right)^{-\frac{2(4+p)}{p}},
\end{eqnarray}
where $\tau \approx \frac{\kappa}{2\chi_0} e^{\frac{2}{\kappa}\chi_0 t}$.
In this limit $t\rightarrow +\infty$, the comoving time is 
$\tau \rightarrow +\infty$ for
$p>0$ while $\tau \rightarrow 0$ for $p<0$.
It is interesting to note that all cases approach the flat Minkowski
space time asymptotically without encountering any divergent curvature.
From the time evolution of the scale factor $\dot{a}(\tau) \sim  sgn(p)$, 
$\ddot{a}(\tau)\sim sgn(p)$, we note that for $p>0$, the universe 
exhibits accelerating expansion and for $p<0$, it does accelerating 
contraction. 

 As a result, the first branch exhibits the acceleration expansion 
for $0< p <8$ with the bounded curvature which is asymptotically flat 
in the far past and future. For $-4< p <0$, the universe
starts at the flat space-time and exhibits the accelerating
contraction and reaches the flat space-time. 

 Let us now analyze the second case, $\Omega_0 =-\chi_0$ ($p>0$). 
From Eqs.~(\ref{sol:Omega}) and (\ref{sol:chi}), the solutions are
given by 
\begin{eqnarray}
e^{-\frac{8}{p}\rho}+\frac{\kappa}{2}
	e^{-\frac{16}{p\kappa}\chi_0 t} \rho &=&0, \\
e^{-2\phi}+\frac{p\kappa}{8}\phi &=& -\chi_0 t. 
\end{eqnarray}
The extremum condition of the scalar curvature yields
\begin{eqnarray}
\dot{R}(t)& =& -\frac{p}{\kappa}\chi_0^3 
e^{-2\phi} \exp\left(\frac{4}{\kappa}e^{-2\phi}\right)\left(e^{-2\phi}
- \frac{p\kappa}{16}\right)^{-5}\left[e^{-4\phi} -
\left(\frac{p\kappa}{16} + \frac{\kappa}{2}\right)e^{-2\phi} - 
\frac{p\kappa^2}{64}\right] \nonumber \\
&=&0
\end{eqnarray}
when $e^{-2\phi}=\frac{\kappa}{32}\left[p+8 \pm \sqrt{p^2+32p+64}\right]$.

For the far past $t\rightarrow -\infty$ $(\tau \rightarrow -\infty)$, 
the metric is asymptotically written as
\begin{eqnarray}
ds^2 &\rightarrow& -\exp\left[ -\frac{4}{\kappa}
	e^{\frac{16}{p\kappa}\chi_0 t}\right]
	(dt^2-dx^2) 	\nonumber \\
	&=& -d\tau^2 + a^2(\tau) dx^2 , 
\end{eqnarray}
where the scale factor is
\begin{equation}
a(\tau) \approx \exp\left[- \frac{2}{\kappa}
\exp\left(\frac{16}{p\kappa}\chi_0\tau\right)\right]
\end{equation}
and $\tau \approx  t$. The curvature scalar asymptotically vanishes 
at $\tau \rightarrow -\infty$,
\begin{equation}
R \rightarrow \left(-\frac{16}{p\kappa}\right)^3 \chi_0^2
\exp\left(\frac{16}{p\kappa}\chi_0 \tau\right).
\end{equation}
It has the maximum value at certain time similarly to the scalar
curvature of the first branch. The time evolutions of the scale 
factor are
\begin{eqnarray}
\dot{a} (\tau) &\sim & sgn(p),\\ 
\ddot{a} (\tau) &\sim & sgn(p).  
\end{eqnarray}
This shows that the scale factor starts at the finite value and increases, and the 
accelerating expansion is possible.   

On the other hand, for $t \rightarrow +\infty$ ($\tau \rightarrow +\infty$),
the metric and the scale factor are
\begin{eqnarray}
ds^2 &\rightarrow& -(-\chi_0 t)^{-\frac{p}{4}}e^{\frac{4\chi_0}{\kappa}t}        (dt^2 -dx^2) \nonumber \\
	&=& -d\tau^2 + a^2(\tau) dx^2 ,\\
a(\tau) &\approx & \frac{2}{\kappa}\chi_0 \tau \left[
-\frac{\kappa}{2}\ln\left(\frac{2}{\kappa}\chi_0\tau\right)
\right]^{-\frac{p}{8}} ,
\end{eqnarray}
where $\tau \approx \frac{\kappa}{2\chi_0}e^{\frac{2}{\kappa}\chi_0 t}$.
The asymptotic curvature is
\begin{equation}
R \rightarrow \frac{p}{4}\left(\frac{2}{\kappa}\tau\right)^{-2}\left[
	-\frac{\kappa}{2}\ln\left(\frac{2}{\kappa}\chi_0\tau\right)
	\right]^{\frac{p-8}{4}}
\end{equation}
and it approaches the flat Minkowskian space-time. As a comment, when 
$p <0$($\tau\rightarrow 0$), the curvature diverges, and we can not 
obtain the bounded curvature on the contrary to the first branch.
For $p>0$, the time evolution of the scale factors are given by
\begin{eqnarray}
\dot{a} (\tau) &\approx& \frac{2}{\kappa}\chi_0  \qquad > 0 , \\
\ddot{a} (\tau)&\approx& 
	\frac{p\kappa}{16}\chi_0\frac{1}{\tau}  \qquad >0 \label{2:d2a2}.
\end{eqnarray}
As a result, the second branch $(p>0)$ exhibits the accelerating
expansion, and the curvature scalar is bounded and asymptotically
flat in the far past and future.

\begin{center}
{\bf IV. de Sitter Universe ($p=8$)}
\end{center}
%\section{de Sitter Universe ($p=8$)}\label{sec:p=8}

 We briefly discuss the case of $p=8$. From the beginning by setting 
$p=8$, the de Sitter space is obtained as
\begin{equation}
a(\tau) \sim e^{-A\Sigma\tau}, \qquad R = {\rm const}.\, ,
\end{equation}
where $A$ and $\Sigma$ are integration constants ($A\Sigma<0$) and
the range of comoving time is $-\infty <\tau<0$. Another solution
(see the Eq.~(\ref{cl:+})) can be given by the flat Minkowskian
space-time for $0<\tau < +\infty$. These two solutions are not
smoothly connected at $\tau =0$. The de Sitter universe of the
constant curvature abruptly changes to the flat space-time.
In the quantized theory, however, the metric is asymptotically de
Sitter type at $t \rightarrow -\infty$ for the first quantum branch,  
\begin{eqnarray}
ds^2 &\rightarrow& - (\chi_0 t)^{-2}(dt^2 - dx^2) \nonumber \\
	&=& -d\tau^2 + e^{-2\chi_0 \tau}dx^2,
\end{eqnarray}
where $\chi_0 t \approx e^{\chi_0\tau}$ and follows the same behavior 
of the first quantum branch at $\tau \rightarrow +\infty$ in sect.~III. 
The quantum corrected maximum scalar curvature is
\begin{equation}
R \rightarrow 2 (\chi_0)^2
\end{equation}
at $\tau \rightarrow -\infty$, and monotonically decreases and
approaches zero. So the universe for the $p=8$ case exhibits 
the accelerating expansion from the de Sitter space-time is smoothly 
connected with the accelerating expansion which ends up with the flat 
Minkowskian space-time in the first branch. There is no de Sitter
phase in the second quantum branch.

\begin{equation}
{\bf V.\ Discussion}
\end{equation}
%\section{Discussion}\label{sec:discuss}

  Especially we comment on the case $p=2$ corresponding to
the s-wave sector of Einstein gravity. Classically, it is not clear 
whether or not the superinflation is possible except for the case of 
the strong coupling constant. However, the nonvanishing cosmological 
constant can be canceled effectively by the induced cosmological 
constant in the quantized theory. So the quantized behaviors are very 
similar to those of the CGHS model~\cite{rey,mr}. Further, we note
that the radius of the two-sphere in four dimensional sense is
proportional to $e^{-2\phi}$, and the two quantum branches are some
kind of T-duality~\cite{gsw} related instead of the scale factor
duality for $p=4$ case. 

 In summary, we have showed that the s-wave sector of the Einstein 
gravity $(p=2)$ in the quantized theory exhibits  
the accelerating expansion and ends up with the small accelerating
expansion ($\ddot{a}>0$)in both branches. 
Especially, for $p=8$, the asymptotic de Sitter space
is smoothly connected with the flat Minkowskian space-time in
the first branch of the quantized theory. 
Further, for $-4<p<0$ in the first branch, the universe is 
accelerating contraction from the Minkowski space-time and has 
no curvature singularity. In this case, the dilaton plays a matter
role and gives an attractive force which is of relevance to the 
contraction of the universe.

\begin{equation}
{\bf Acknowledgments}
\end{equation} 
%\section*{Acknowledgments}\label{sec:ackno}

This work was supported by Ministry of Education, 1996, Project No. 
BSRI-96-2414, and Korea Science and 
Engineering Foundation through the Center for Theoretical Physics
in Seoul National University(1997).

%%%%%%%%%%%%%%%%%%%% References %%%%%%%%%%%%%%%%%%%%%%%%%

\end{document}